\documentclass[12pt,a4paper]{conference}

\usepackage{fancyhdr}
\usepackage{graphicx,amsmath,amssymb,cite}
\usepackage{multind}
\makeindex{author} \makeindex{subject}
\newcommand{\beq}{\begin{equation}}
\newcommand{\eeq}[1]{\label{#1}\end{equation}}
\newcommand{\eeqn}{\end{equation}}

\newcommand{\beqa}{\begin{eqnarray}}
\newcommand{\eeqa}[1]{\label{#1}\end{eqnarray}}
\newcommand{\eeqan}{\end{eqnarray}}

\let\bar=\overbar

\newcommand{\Dslash}{\not{\hbox{\kern-4pt $D$}}}
\newcommand{\dslash}{\not{\hbox{\kern-2pt $\del$}}}

\newcommand{\msb}{{\bar{\ssstyle M \kern -1pt S}}}

\begin{document}
%%%%%%%%%%%%%%%%%%%%%%%%%%%%%%%%%%%%%%%%%%%%%%%%%%%%%%%%%%%%%%%%%%%%%%%

\Chapter{Studying pion effects in the quark propagator}
           {Studying pion effects in the quark propagator}{D. Nickel \it{et al.}}
\vspace{-5 cm}\includegraphics[width=6 cm]{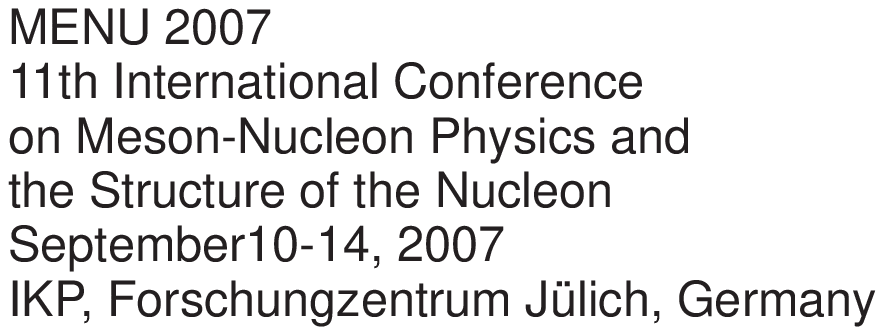}
%\bigskip\bigskip
\vspace{4 cm}

\addcontentsline{toc}{chapter}{{\it D. Nickel}} \label{authorStart}
%%%%%%%%%%%%%%%%%%%%%%%%%%%% NEW SWITCHES %%%%%%%%%%%%%%%%%%%%%%%%%%%%%%

\begin{raggedright}

%%%%%%%%%%%%%%%%
%%% The authors
%%%%%%%%%%%%%%%%

D. Nickel$^{\star}$$^,$\footnote{E-mail address:
dominik.nickel@physik.tu-darmstadt.de},
C.S. Fischer$^{\star}$,
J. Wambach$^{\star}$
\bigskip\bigskip

%%%%%%%%%%%%%
%%% addresses
%%%%%%%%%%%%%

$^{\star}$Institut f\"ur Kernphysik, Technische Universit\"at
Darmstadt, Germany

\end{raggedright}

\begin{center}
\textbf{Abstract}
\end{center}
Within the framework of Schwinger-Dyson and 
Bethe-Salpeter equations we investigate the importance of pions for
the quark-gluon interaction.
To this end we choose a truncation for the
quark-gluon vertex that includes intermediate pion degrees of freedom and
adjust the interaction such that unquenched lattice results for various current
quark masses are reproduced.
The corresponding Bethe-Salpeter kernel is constructed from
constraints by chiral symmetry.
After extrapolation to the physical point we find a considerable
contribution of the pion back reaction to the quark mass function as well as
to the chiral condensate. The quark wave function is less affected.

\section{Introduction and framework}
Dynamical chiral symmetry breaking is one of the most striking properties of
low-energy QCD. The resulting appearance of massless pions in the
chiral limit is the basis of a systematic expansion for hadronic
observables in terms of chiral perturbation theory.
The difference between quenched and unquenched simulations in lattice QCD is
also attributed to the importance of the pions' dynamics.
Here we study the pion contribution and back reaction on the quark degrees of
freedom in the Green's function approach  to Landau gauge QCD using
Schwinger-Dyson and Bethe-Salpeter equations
(SDE/BSE)~\cite{Maris:1997hd,Alkofer:2000wg}.
We consider this as an extension of previous NJL-model
studies~\cite{Dmitrasinovic:1995cb} to full QCD
and to be complementary to previous studies within the SDE/BSE approach
neglecting the back reaction of
pions~\cite{Watson:2004jq,Fischer:2003rp}.

Our starting point is the SDE of the quark propagator
\begin{eqnarray}
  S^{-1}(p) &=& Z_{2}S^{-1}_{0}(p)+\Sigma(p)\,, \label{DSE1}
\end{eqnarray}
where $S^{-1}_{0}(p) = i p\!\cdot\!\gamma + m$ denotes the inverse bare
quark-propagator,  while $S^{-1}(p) = (i p\cdot\gamma  +
M(p^2))/Z_{f}(p^2)$ is the dressed propagator being parameterized by the quark
mass $M(p^2)$ and the quark wave function $Z_f(p^2)$. $Z_{2}$ is the
renormalization factor of the quark field.
The quark self energy in Landau gauge is given by 
\begin{eqnarray}
  \Sigma(p) &=& g^{2}C_{F}Z_{1F}\int\!\!\frac{d^4q}{(2 \pi)^4}\,
  \gamma_{\mu}S(q)\Gamma_{\nu}(q,k)
  \,
  \frac{Z(k^2)}{k^2}
  \left(\delta_{\mu \nu} - \frac{k_\mu k_\nu}{k^2}\right) 
  \,,
  \label{fullSigma}
\end{eqnarray}
with $k=p-q$, the Casimir $C_{F}=(N_{c}^{2}-1)/(2N_{c})$ and the
renormalization  factor $Z_{1F}$ of the quark gluon vertex. The self energy
depends on the fully dressed quark-gluon vertex $\Gamma_{\nu}(q,k)$ and the
gluon dressing function $Z(k^2)$.

\begin{figure}
\centerline{\includegraphics[width=14cm]{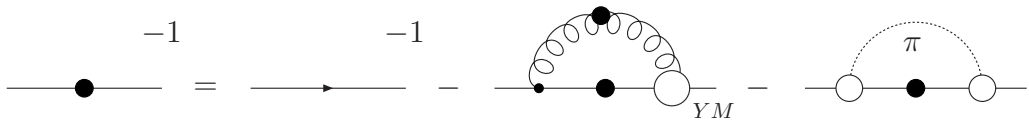}}
\caption{The approximated Schwinger-Dyson equation for the quark propagator
with effective one-gluon exchange and one-pion exchange. 
\label{fig:quarkdse2}}
\vspace{-1mm}
\end{figure}

The widely used rainbow-ladder approximation amounts to the replacement
\begin{eqnarray}
\gamma_{\mu} Z(k^2) \Gamma_{\nu}(q,k) \rightarrow \gamma_{\mu} \Gamma(k^2) 
                                                  \gamma_{\nu}\,, \label{RL}
\end{eqnarray}
where $\Gamma(k^2)$ can be viewed as a combination of the gluon dressing 
function and a purely $k^2$-dependent dressing of the $\gamma_\nu$-part of the
quark-gluon vertex.
%%%%%%%%%%%%%%%%%%%%%%%%%%%%%%%%%%%%%%%%%%%%%%%%%%%%%%%%%%%%%%%%%%%%%%%%%%%%
\begin{figure}
\centerline{\includegraphics[width=7cm]{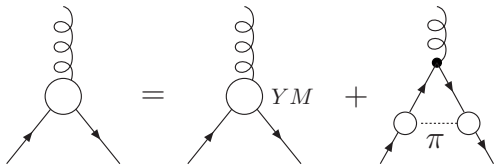}}
\caption{The approximated Schwinger-Dyson equation for the quark-gluon
vertex. All internal propagators are fully dressed. \label{fig:Vertexdse2}}
\vspace{-3mm}
\end{figure}
%%%%%%%%%%%%%%%%%%%%%%%%%%%%%%%%%%%%%%%%%%%%%%%%%%%%%%%%%%%%%%%%%%%%%%%%%%%%
Aiming at an extension of this approximation which includes explicite pion
degrees of freedom, we can motivate the quark-gluon vertex diagrammatically
shown in Fig.~\ref{fig:Vertexdse2} by its SDE (see Ref.~\cite{Fischer:2007ze}
for details). The main idea is to single out the leading contribution involving
pions and approximate the remaining part as the vertex used in the
rainbow-ladder approximation.
Using this ansatz in the SDE of the quark propagator we can motivate - after
intermediate steps - the truncation diagrammatically shown in
Fig.~\ref{fig:quarkdse2}.
%%%%%%%%%%%%%%%%%%%%%%%%%%%%%%%%%%%%%%%%%%%%%%%%%%%%%%%%%%%%%%%%%%%%%%%%%%%%
%%%%%%%%%%%%%%%%%%%%%%%%%%%%%%%%%%%%%%%%%%%%%%%%%%%%%%%%%%%%%%%%%%%%%%%%%%%%%
Guided by chiral symmetry constraints we construct a Bethe-Salpeter
kernel that guarantees the pion to be massless in the chiral limit (see again
Ref.~\cite{Fischer:2007ze} for details).

\section{Results}
%%%%%%%%%%%%%%%%%%%%%%%%%%%%%%%%%%%%%%%%%%%%%%%%%%%%%%%%%%%%%%%%%%%%%%%%%%%%
\begin{figure}
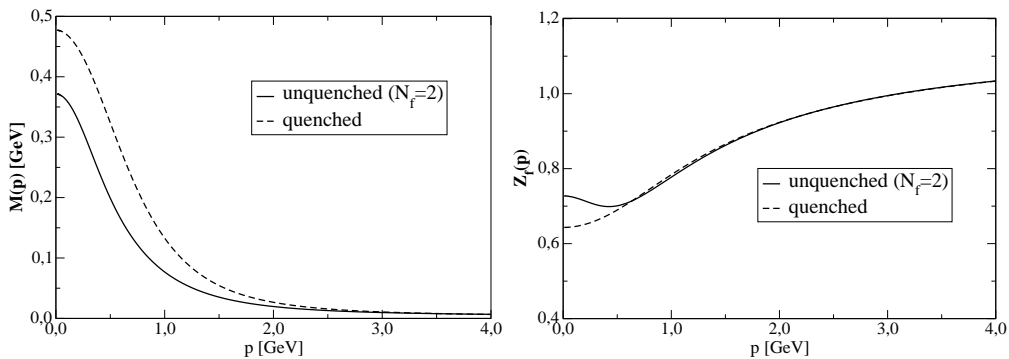

\centerline{\includegraphics[width=6.5cm]{PI_M.eps}
      \includegraphics[width=6.5cm]{PI_Z.eps}}
\caption{The quenched and unquenched
  ($N_f$=2) quark mass (left) and wave function (right) for
  physical up/down quarks with $M(2.9 \,\mbox{GeV})=10\,\mbox{MeV}$.}
\label{fig:quark}
\vspace{-6mm}
\end{figure}
%%%%%%%%%%%%%%%%%%%%%%%%%%%%%%%%%%%%%%%%%%%%%%%%%%%%%%%%%%%%%%%%%%%%%%%%%%%%%
Given the unquenched lattice QCD results for the quark propagator in Landau
gauge at relatively large quark masses~\cite{Bowman:2005vx} and a
parameterization for the dressing of the quark-gluon vertex in rainbow-ladder
approximation adopted from similar investigations of quenched lattice QCD
results~\cite{Bhagwat:2003vw}, we can adjust the parameters of
the rainbow-ladder contribution to our truncation of the quark-gluon vertex.
We get nice agreement
with the lattice QCD results for the quark mass function as well as for the
wave function (see Ref.~\cite{Fischer:2007ze}).
Switching off the pion contribution to the quark self-energy and furthermore
comparing to unquenched lattice QCD results, we find the pion contribution to
be overestimated.
This may be due to a further numerical approximation and is detailed in
Ref.~\cite{Fischer:2007ze}. 

%%%%%%%%%%%%%%%%%%%%%%%%%%%%%%%%%%%%%%%%%%%%%%%%%%%%%%%%%%%%%%%%%%%%%%%%%%%%
\begin{figure}
\centerline{\includegraphics[width=7.5cm]{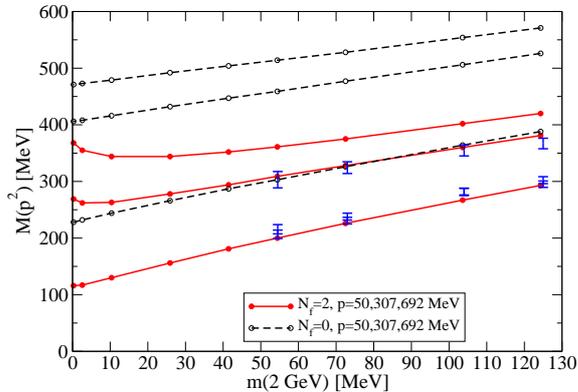}}
\caption{Chiral extrapolation of the mass function at stated momenta
  from bottom to top including the pion effects in the quark-gluon
  vertex (red) and neglecting it (black). Also given are lattice QCD
  results for the momenta corresponding to the touching red lines.}
\label{fig:extrap}
\vspace{-5mm}
\end{figure}
%%%%%%%%%%%%%%%%%%%%%%%%%%%%%%%%%%%%%%%%%%%%%%%%%%%%%%%%%%%%%%%%%%%%%%%%%%%%%

The general tendency is however in line with the lattice QCD results and we
can perform an extrapolation to physical value of the pion mass, {\it i.e.}
current quark mass, and to the chiral limit.
In Fig.~\ref{fig:quark} we also show the quark mass function and the wave
function at the physical point. At small momenta the current quark mass
dependence is most pronounced for the wave function, whereas the dependence of
the mass function is strongest at intermediate momenta.
To illustrate this we present the chiral extrapolation of the mass function at
some momenta accessed in the lattice QCD simulations in Fig.~\ref{fig:extrap}.
In contrast to the rainbow-ladder truncation, which gives an almost linear
extrapolation, we find a sizeable non-linear behavior at small quark masses
and momenta due to the pion contribution.

This work has been supported by the Helmholtz-University Young
Investigator Grant VH-NG-332.

\end{document}